\documentclass[aps,pra,preprint,groupedaddress,showpacs]{revtex4}
\usepackage{bm}
\begin{document}
\preprint{Version 4.0}

\title{Relativistic corrections to the long range interaction between 
closed shell atoms}

\author{Krzysztof Pachucki}
\email[]{krp@fuw.edu.pl} \homepage[]{www.fuw.edu.pl/~krp}

\affiliation{Institute of Theoretical Physics, Warsaw University,
             Ho\.{z}a 69, 00-681 Warsaw, Poland}

\date{\today}

\begin{abstract}
The complete $O(\alpha^2)$ correction to the long range
interaction between neutral closed shell atoms is obtained,
the relation to the asymptotic expansion of the known short range 
interaction at the atomic scale is presented and a general interaction potential
which is valid in the whole range of the inter atomic distances is constructed.
\end{abstract}


\maketitle
The retarded long-range interaction between neutral systems
was first considered by Casimir and Polder in their pioneering 
work in \cite{casimir}. At large distances atoms interact predominantly by 
the two-photon exchange with the nonrelativistic dipole interaction 
$-e\,\vec r\cdot\vec E$. There were various extensions of their result.
First of all, Feinberg and Sucher in \cite{general} (see also a longer review
in \cite{report}) expressed the two-photon exchange interaction 
(dispersion forces) in terms of the well-defined physical
quantities, the invariant amplitudes which
describe the elastic scattering of a photon by an atom. These
amplitudes are often called dynamic polarizabilities.
In this way the authors obtained a formally exact result, 
as all relativistic corrections are included in the dynamic polarizabilities.
In the nonrelativistic limit their result reduces to that of Casimir and Polder.
In the limit of large distances their result
is expressible in terms of a static electric and magnetic polarizabilities only,
similarly to the nonrelativistic Casimir-Polder interaction.
The calculation of relativistic corrections to the dynamic polarizability
is not a simple problem. Therefore in later works a different approach 
to dispersion forces was developed which based on a reformulation of nonrelativistic 
quantum electrodynamics. Apart from the rederivation of
the Casimir-Polder result, there were derived closed formulae for higher multipole 
interactions such as electric quadrupole
\cite{equad1, equad2, equad3} and electric octupole \cite{eoctu} for isotropic systems, 
magnetic dipole  and diamagnetic couplings for chiral
molecules \cite{equad2}.  In this work, using a different reformulation
of nonrelativistic quantum electrodynamics \cite{lwqed}, we present a
systematic derivation of all $O(\alpha^2)$ contributions to the Casimir-Polder 
potential of closed shell atoms, including corrections which have
not been considered so far. The obtained result is expressible in terms
of various corrections to the nonrelativistic dynamic polarizability,
in agreement with the general result of Feinberg and Sucher in \cite{general}. 
Moreover, we present the relation of relativistic corrections
to the Casimir-Polder interaction to the short range
nonrelativistic expansion 
and construct an interaction potential which is valid in 
the whole range of inter-atomic distances.

Let us first define the $\alpha$ expansion of the interaction potential
in the clamped nuclei approximation. We use natural units, where
$c=\hbar=\epsilon_0=1$ and denote by $m$ the electron mass.
The total energy of a system consisting of two neutral atoms
is a function of the fine structure constant $\alpha$ and 
the distance $R$ between these atoms
\begin{equation}
E = E(\alpha,m\,R)\,. \label{01}
\end{equation}
The nonrelativistic expansion in $\alpha$ depends on the magnitude of $R$.
According to quantum electrodynamics, if $R$ is of the order of an atomic 
size $R\sim 1/(m\,\alpha)$, then
this expansion at constant $m\,\alpha\,R$ takes the form
\begin{equation}
E(\alpha,m\,R) = E^{(2)}(m\,\alpha\,R) + E^{(4)}(m\,\alpha\,R) +
E^{(5)}(m\,\alpha\,R) + O(\alpha^6)\,,\label{02}
\end{equation}
where $E^{(2)}$ is the nonrelativistic energy of order $m\,\alpha^2$
of a systems of two atoms including Coulomb interactions between all electrons, 
$E^{(4)}$ is the leading relativistic correction of order $m\,\alpha^4$, 
which is given by the Breit-Pauli Hamiltonian $\delta H$ \cite{bs}. 
We include below only the terms which do not vanish for closed shell atoms
\begin{eqnarray}
E^{(4)} &=& \langle \delta H \rangle\,, \label{03}\\
\delta H &=& \sum_a \biggl[-\frac{\vec p^{\,4}_a}{8\,m^3} +
\frac{\pi\,Z\,\alpha}{2\,m^2}\,\delta^3(r_a)\biggr] 
+\sum_{a>b} \biggl[
\frac{\pi\,\alpha}{m^2}\, \delta^3(r_{ab})
-\frac{\alpha}{2\,m^2}\,p_a^i\,
\biggl(\frac{\delta^{ij}}{r_{ab}}+\frac{r^i_{ab}\,r^j_{ab}}{r^3_{ab}}
\biggr)\,p_b^j\biggr]\,,\label{04}
\end{eqnarray}
where the sum goes over all electrons of both atoms.
$E^{(5)}$ is the QED correction of order $m\,\alpha^5$.
It consists of various terms, among others the Araki-Sucher term \cite{helamb1,
helamb2, helamb3}, which is dominating at large atomic distances,
\begin{eqnarray}
E^{(5)} &=& \sum_{a>b} -\frac{14}{3}\,m\,\alpha^5\,\biggl\langle
\frac{1}{4\,\pi}\,P\biggl(\frac{1}{(m\,\alpha\,r_{ab})^3}\biggr)\biggr\rangle
+\ldots \label{05}
\end{eqnarray}

On the other hand, if $R$ is of the order of atomic transition wavelength,
namely $R\sim 1/(m\,\alpha^2)$  then the $\alpha$ expansion at constant
$m\,\alpha^2\,R$
takes a completely different form
\begin{equation}
E(\alpha,m\,R) = E_{\rm free}(\alpha) +  
E_{CP}(m\,\alpha^2 R) + \delta^{(2)} E_{CP}(m\,\alpha^2 R) + \ldots \label{06}
\end{equation}
where $E_{\rm free}$ is the energy of separate atoms,
$E_{CP}$ is a Casimir-Polder potential \cite{casimir} and 
$\delta^{(2)} E_{CP}$ is the leading $\alpha^2$ 
relativistic correction, which is the subject of this work.
The form of this expansion results from the long wavelength
formulation of quantum electrodynamics, see Ref. \cite{lwqed}. 
This relativistic correction to the interaction energy at large atomic distances
is obtained from the effective interaction $H_I$
of an atom with the slowly varying electromagnetic field \cite{lwqed}
\begin{eqnarray}
H_I &=& \sum_a  -e\,\vec r_a\cdot \vec E-\frac{e}{2}\,\biggl(r_a^i\,r_a^j-
\frac{\delta^{ij}}{3}\,r_a^2\biggr)\,E^i_{,j}-
\frac{e}{30}\,r_a^2\,r_a^i\,E^i_{,jj} \nonumber \\ &&
-\frac{e}{6\,m}\,\bigl(L_a^i\,r_a^j+r_a^j\,L_a^i\bigr)\,B^i_{,j}+
\frac{e^2}{8\,m^2}\,\bigl(\vec r_a\times\vec B\bigr)^2\,, \label{07}
\end{eqnarray}
where $\vec L = \vec r\times\vec p$, $\vec E = \vec E(0)$, $\vec B = \vec
B(0)$ are fields at the position of nucleus,
and spin dependent terms have been neglected as we consider only closed shell
atoms. The sum in Eq. (\ref{07}) goes over all electrons of one atom.
For simplicity, we will assume this sum is present implicitly in all 
the formulas below. 
The leading Casimir-Polder interaction comes from 
the two-photon exchange with the interaction 
$-e\,\vec r_a\cdot \vec E$.  Using the temporal gauge for the photon propagator
$A^0 = 0$, it is \cite{gauge}
\begin{equation}
E_{CP} = -\frac{e^4}{2}\,\int_{-\infty}^\infty\,\frac{d\omega}{2\,\pi\,i}\,
\alpha_{EA}^{ik}\,\alpha_{EB}^{jl}\,g^{ij}\,g^{kl}\,, \label{08}
\end{equation}
where
\begin{eqnarray}
g^{ij} &=& \int\frac{d^3k}{(2\,\pi)^3}\,e^{i\,\vec k\cdot\vec R}\,
\frac{(\omega^2\,\delta^{ik}-k^i\,k^k)}{\omega^2-k^2}\,, \label{09}
\\
\alpha_E^{ij} &=& -\biggl\langle r^i\frac{1}{E-H+\omega}\,r^j +
r^j\frac{1}{E-H-\omega}\,r^i \biggr\rangle\,, \label{10}
\end{eqnarray}
and $\omega$-integration is assumed along the Feynman contour.
This integration contour is deformed to imaginary axis by the replacement
$\omega = i\,\lambda$. The $k$-integral leads to
\begin{equation}
g^{ij} = (\lambda^2\,\delta^{ij}-\partial^i\,\partial^j)\,\frac{e^{-\lambda\,R}}{R}
= \frac{e^{-\lambda\,R}}{R^3}\biggl[
\delta^{ij}(\lambda^2\,R^2+\lambda\,R+1)-\frac{R^i\,R^j}{R^2}\,
(\lambda^2\,R^2+3\,\lambda\,R+3)\biggr]\,. \label{11}
\end{equation}
For spherically symmetric states $\alpha_X^{ij} = \delta^{ij}\,\alpha_X$ and
\begin{eqnarray}
E_{CP} &=&-\frac{4\,\alpha^2}{9\,\pi}\,\int_0^\infty\,d\lambda
\biggl\langle\vec r\frac{H-E}{(H-E)^2+\lambda^2}\,\vec r\biggr\rangle_A\,
\biggl\langle\vec r\frac{H-E}{(H-E)^2+\lambda^2}\,\vec r\biggr\rangle_B
\nonumber \\ &&\times
\frac{\lambda^4\,e^{-2\,\lambda\,R}}{R^2}\,
\biggl(1+\frac{2}{\lambda\,R}+\frac{5}{(\lambda\,R)^2}+
\frac{6}{(\lambda\,R)^3}+\frac{3}{(\lambda\,R)^4}\biggr)\label{13} \\ &=&
-\frac{\alpha^2}{\pi}\,\int_0^\infty\,d\lambda\,\alpha_{EA}(i\,\lambda)\,
\alpha_{EB}(i\,\lambda)\,\frac{\lambda^4\,e^{-2\,\lambda\,R}}{R^2}\,
\biggl(1+\frac{2}{\lambda\,R}+\frac{5}{(\lambda\,R)^2}+
\frac{6}{(\lambda\,R)^3}+\frac{3}{(\lambda\,R)^4}\biggr)\,. \nonumber
\end{eqnarray}
This result has been obtained by Casimir and Polder in \cite{casimir}. 
We consider here the $\alpha^2$ correction and represent it as a sum of five terms
\begin{equation}
\delta^{(2)} E_{CP} = \delta^{(2)}_0 E_{CP} +  \delta^{(2)}_1 E_{CP} + 
 \delta^{(2)}_2 E_{CP} +  \delta^{(2)}_3 E_{CP} +  \delta^{(2)}_4 E_{CP}\,. \label{14}
\end{equation} 
$\delta^{(2)}_0 E_{CP}$ is due to the Breit-Pauli correction to $H$, $E$
and state $\phi$ in Eq. (\ref{13}). For simplicity we consider corrections 
only to the atom $A$, therefore only one matrix element in Eq. (\ref{13}) 
is to be modified in $\delta^{(2)}_0 E_{CP}$, according to
\begin{eqnarray}
\delta^{(2)}_0 E_{CP} &=&-\frac{4\,\alpha^2}{9\,\pi}\,\int_0^\infty\,d\lambda\;
\delta \biggl\langle\vec r\frac{H-E}{(H-E)^2+\lambda^2}\,\vec r\biggr\rangle_A\,
\biggl\langle\vec r\frac{H-E}{(H-E)^2+\lambda^2}\,\vec r\biggr\rangle_B
\nonumber \\ &&\times
\frac{\lambda^4\,e^{-2\,\lambda\,R}}{R^2}\,
\biggl(1+\frac{2}{\lambda\,R}+\frac{5}{(\lambda\,R)^2}+
\frac{6}{(\lambda\,R)^3}+\frac{3}{(\lambda\,R)^4}\biggr)\,, \label{15}\\
\delta\biggl\langle\vec r\frac{H-E}{(H-E)^2+\lambda^2}\,\vec
r\biggr\rangle &\equiv& 
2\,\biggl\langle\delta H\frac{1}{(E-H)'}\,\vec r\frac{H-E}{(H-E)^2+\lambda^2}\,\vec
r\biggr\rangle\nonumber \nonumber \\&&
+\frac{1}{2}\,\biggl\langle\vec r\,\frac{1}{H-E+i\,\lambda}\,(\langle\delta H\rangle-\delta H)\,
\frac{1}{H-E+i\,\lambda}\vec r\biggr\rangle \nonumber \\&&
+\frac{1}{2}\,\biggl\langle\vec r\,\frac{1}{H-E-i\,\lambda}\,(\langle\delta H\rangle-\delta H)\,
\frac{1}{H-E-i\,\lambda}\vec r\biggr\rangle\,. \label{16}
\end{eqnarray} 
This correction has recently been considered in Ref. \cite{moszynski}. 
All remaining corrections to the Casimir-Polder interaction energy are
obtained by modification of a dipole interaction $-e\,\vec r\cdot\vec E$ 
by various couplings as given by Eq. (\ref{07}).  $\delta^{(2)}_1 E_{CP}$ 
comes from the quadrupole term 
$ -e/2\,(r_a^i\,r_a^j-\delta^{ij}\,r_a^2/3)\,E^i_{,j}$ in Eq. (\ref{07}),
\begin{eqnarray}
\delta^{(2)}_1 E_{CP} &=& -\frac{\alpha^2}{120}\,\int_0^\infty\,
\frac{d\lambda}{2\,\pi}
\biggl\langle r^{mn}\,\frac{2\,(H-E)}{(H-E)^2+\lambda^2}\,r^{mn}\biggr\rangle_A\,
\biggl\langle r^l\,\frac{2\,(H-E)}{(H-E)^2+\lambda^2}\,r^l\biggr\rangle_B
\nonumber \\ && 
(\partial^jg^{ik}\,\partial^ig^{jk} + \partial^pg^{ik}\,\partial^pg^{ik})\,, \label{17}
\end{eqnarray}
where 
\begin{equation}
r^{ij} \equiv r^i\,r^j-\frac{\delta^{ij}}{3}\,r^2\,. \label{18}
\end{equation}
After contracting $i,j,k$ indices it becomes
\begin{eqnarray}
\delta^{(2)}_1 E_{CP} &=& -\frac{\alpha^2}{30\,\pi}\,\int_0^\infty\,d\lambda\,
\biggl\langle r^{mn}\,\frac{2\,(H-E)}{(H-E)^2+\lambda^2}\,r^{mn}\biggr\rangle_A\,
\biggl\langle r^l\,\frac{2\,(H-E)}{(H-E)^2+\lambda^2}\,r^l\biggr\rangle_B
\nonumber \\ && 
\frac{\lambda^6}{R^2}\,e^{-2\,\lambda\,R}\,
\biggl(1+\frac{6}{\lambda\,R}+\frac{27}{(\lambda\,R)^2}+
\frac{84}{(\lambda\,R)^3} + \frac{162}{(\lambda\,R)^4}+
\frac{180}{(\lambda\,R)^5}+\frac{90}{(\lambda\,R)^6}\biggr)\,. \label{19}
\end{eqnarray}
This result was first obtained by Jenkins, Salam and Thirunamachandran
in \cite{equad2}. $\delta^{(2)}_2 E_{CP}$ is the correction that comes from the term 
$-e/30\;r_a^2\,r_a^i\,E^i_{,jj}$ in Eq. (\ref{07}). It is very similar to the 
dipole-dipole interaction and can easily be obtained 
on the basis of Eq. (\ref{13})
\begin{eqnarray}
\delta^{(2)}_2 E_{CP} &=&-\frac{4\,\alpha^2}{135\,\pi}\,\int_0^\infty\,d\lambda
\biggl\langle \vec r\frac{H-E}{(H-E)^2+\lambda^2}\,r^2\,\vec r\biggr\rangle_A\,
\biggl\langle\vec r\frac{H-E}{(H-E)^2+\lambda^2}\,\vec r\biggr\rangle_B
\nonumber \\ &&\times
\frac{\lambda^6\,e^{-2\,\lambda\,R}}{R^2}\,
\biggl(1+\frac{2}{\lambda\,R}+\frac{5}{(\lambda\,R)^2}+
\frac{6}{(\lambda\,R)^3}+\frac{3}{(\lambda\,R)^4}\biggr)\,. \label{20}
\end{eqnarray}
$\delta^{(2)}_3 E_{CP}$ is due to
$-e/(6\,m)\,\bigl(L^i_a\,r^j_a+r^j_a\,L^i_a\bigr)\,B^i_{,j}$
and can be regarded as another correction to the electric dipole coupling 
\begin{eqnarray}
\delta^{(2)}_3 E_{CP} &=&-\frac{4\,\alpha^2}{27\,\pi}\,\int_0^\infty\,d\lambda
\biggl\langle\vec r\frac{1}{(H-E)^2+\lambda^2}\,\biggl(-\frac{i}{2}\biggr)
\,\bigl(\vec L\times\vec r-\vec r\times\vec L\bigr)\biggr\rangle_A\,
\biggl\langle\vec r\frac{H-E}{(H-E)^2+\lambda^2}\,\vec r\biggr\rangle_B
\nonumber \\ &&\times
\frac{\lambda^6\,e^{-2\,\lambda\,R}}{R^2}\,
\biggl(1+\frac{2}{\lambda\,R}+\frac{5}{(\lambda\,R)^2}+
\frac{6}{(\lambda\,R)^3}+\frac{3}{(\lambda\,R)^4}\biggr)\,. \label{21}
\end{eqnarray} 
The last correction $\delta^{(2)}_4 E_{CP}$ is due to 
$e^2/(8\,m^2)\,\bigl(\vec r_a\times\vec B\bigr)^2$ and reads
\begin{eqnarray}
\delta^{(2)}_4 E_{CP} &=& -\frac{\alpha^2}{9\,\pi}\,\int_0^\infty 
d\lambda\,\langle r^2\rangle_A\,
\biggl\langle\vec r\frac{H-E}{(H-E)^2+\lambda^2}\,\vec r\biggr\rangle_B\,
\frac{\lambda^4}{R^2}\,e^{-2\,\lambda\,R}\,
\biggl(1+\frac{2}{\lambda\,R} + \frac{1}{(\lambda\,R)^2} \biggr)\,. \label{22}
\end{eqnarray}
The complete $O(\alpha^2)$ correction is a sum of Eqs. 
(\ref{15},\ref{19},\ref{20},\ref{21},\ref{22}) as given
by Eq. (\ref{14}). Here, $\delta_1^{(2)} E_{CP}$ comes from
interaction between the electric dipole and the electric quadrupole polarizabilities, 
$\delta_1^{(4)} E_{CP}$ is the interaction energy of the electric dipole polarizability
with the magnetic susceptibility, and  
$E_{CP}+\delta_0^{(2)} E_{CP}+\delta_2^{(2)} E_{CP}+\delta_3^{(2)} E_{CP}$
is the interaction energy between electric dipole polarizabilities $\alpha_E$
with the relativistic correction $\delta^{(2)} \alpha_E$
\begin{eqnarray}
\alpha_E(i\,\lambda) &\equiv& \frac{2}{3}\left\langle\vec r\,
\frac{H-E}{(H-E)^2+\lambda^2}\,\vec r\right\rangle\,,\\
\delta^{(2)} \alpha_E(i\,\lambda) &=& 
\frac{2}{3}\,\delta \left\langle\vec r\,
\frac{H-E}{(H-E)^2+\lambda^2}\,\vec r\right\rangle+
\frac{2\,\lambda^2}{45}\,\left\langle\vec r\,
\frac{H-E}{(H-E)^2+\lambda^2}\,r^2\,\vec r\right\rangle\nonumber \\ &&
+\frac{2\,\lambda^2}{9}\,\biggl\langle\vec r\frac{1}{(H-E)^2+\lambda^2}
\,\biggl(-\frac{i}{2}\biggr)
\,\bigl(\vec L\times\vec r-\vec r\times\vec L\bigr)\biggr\rangle\,.
\label{dae}
\end{eqnarray}  
We have not found in the literature the complete formula
for the leading relativistic correction to the electric dipole polarizability
of the closed shell atoms as that in the Eq. (\ref{dae}).

Let us now consider the large and small $R$ limit of the interaction energy.
At large $R$, the $\delta^{(0)}_0 E_{CP}$ and $\delta^{(2)}_4 E_{CP}$ contribute
to the $1/R^7$ coefficient, but it is only a small correction on the top of $E_{CP}$. 
Much more interesting is a small $R$ expansion of $E_{CP}$ and $\delta^{(2)} E_{CP}$
and its relation to the large $R$ expansion of energy as a function of $\alpha$ 
and $m\,\alpha\,R$. This relation has been first
considered by Meath and Hirschfelder in \cite{meath}.
The large $R$ expansion of $E^{(i)}(m\,\alpha\,R)$ from Eq. (\ref{02}) reads
\begin{eqnarray}
E^{(2)} &=& E_{\rm free}^{(2)}-m\,\alpha^2\,\biggl[\frac{C_6^{(2)}}{(m\,\alpha\,R)^6} + 
\frac{C_8^{(2)}}{(m\,\alpha\,R)^8}+\ldots\biggr]\,,\label{23}\\
E^{(4)} &=& E_{\rm free}^{(4)}-m\,\alpha^4\,\biggl[\frac{C_4^{(4)}}{(m\,\alpha\,R)^4} + 
\frac{C_6^{(4)}}{(m\,\alpha\,R)^6}+\ldots\biggr]\,,\label{24}\\
E^{(5)} &=& E_{\rm free}^{(5)}-m\,\alpha^5\,\biggl[\frac{C_3^{(5)}}{(m\,\alpha\,R)^3} + 
\frac{C_5^{(5)}}{(m\,\alpha\,R)^5}+\ldots\biggr]\,,\label{25}
\end{eqnarray} 
where $C_i^{(j)}$ are dimensionless constants. The relation to the 
small $R$ expansion of $E_{CP}(m\,\alpha^2 R)$ and $\delta^{(2)}E_{CP}(m\,\alpha^2 R)$
from Eq. (\ref{06}) is the following
\begin{eqnarray}
E_{CP} &=& -m\,\alpha^8\,\biggl[\frac{C_6^{(2)}}{(m\,\alpha^2\,R)^6}
+\frac{C_4^{(4)}}{(m\,\alpha^2\,R)^4}
+\frac{C_3^{(5)}}{(m\,\alpha^2\,R)^3}+\ldots\biggr]\,,\label{26}\\
\delta^{(2)} E_{CP} &=& -m\,\alpha^{10}\,\biggl[
\frac{C_8^{(2)}}{(m\,\alpha^2\,R)^8}
+\frac{C_6^{(4)}}{(m\,\alpha^2\,R)^6}
+\frac{C_5^{(5)}}{(m\,\alpha^2\,R)^5}+\ldots\biggr]\,.\label{27}
\end{eqnarray}
Since both expansions of $E(\alpha,m\,R)$ involve the same coefficients
$C^{(i)}_j$, one can write the general formula
\begin{equation}
E(\alpha,m\,R) = E_{\rm free}(\alpha) 
- \sum_{i,j} m\,\alpha^i\,\frac{C^{(i)}_j}{(m\,\alpha\,R)^j}\,.
\end{equation}
We have checked this by equivalence of $C_i^{(j)}$
coefficients as obtained from these two different expansions,
and they are equal to (in atomic units)
\begin{eqnarray}
C_6^{(2)} &=& \frac{2}{3}\,\biggl\langle
              r_A^i\,r_B^j\,\frac{1}{H_A+H_B-E_A-E_B}\,r_A^i\,r_B^j\biggr\rangle
              \,,\label{28}\\
C_4^{(4)} &=& \frac{2}{9}\,\biggl\langle
              r_A^i\,r_B^j\,\frac{1}{H_A+H_B-E_A-E_B}\,p_A^i\,p_B^j\biggr\rangle
              \,,\label{29}\\
C_3^{(5)} &=& \frac{7}{6\,\pi}\,N_A\,N_B\,,\label{30}\\
C_8^{(2)} &=& \frac{3}{2}\,\biggl\langle 
              \biggl(r_A^i\,r_A^j-\frac{\delta^{ij}}{3}\,r_A^2\biggr)\,r_B^k\,
              \frac{1}{H_A+H_B-E_A-E_B}\,
              \biggl(r_A^i\,r_A^j-\frac{\delta^{ij}}{3}\,r_A^2\biggr)\,r_B^k\biggr\rangle 
              \,,\label{31}\\
C_6^{(4)} &=& \frac{2}{3}\,\biggl\langle
              r_A^i\,r_B^j\,\frac{1}{H_A+H_B-E_A-E_B}\,
              (\langle\delta H_A\rangle-\delta H_A)
              \,\frac{1}{H_A+H_B-E_A-E_B}\,r_A^i\,r_B^j\biggr\rangle\nonumber
              \\ && +
              \frac{4}{3}\,\biggl\langle\delta H_A\,\frac{1}{(E_A-H_A)'}\,
              r_A^i\,\,r_B^j\,\frac{1}{H_A+H_B-E_A-E_B}\,r_A^i\,r_B^j\biggr\rangle
              \nonumber\\ && +
              \frac{3}{5}\,\biggl\langle 
              \biggl(r_A^i\,r_A^j-\frac{\delta^{ij}}{3}\,r_A^2\biggr)\,r_B^k\,
              \frac{1}{H_A+H_B-E_A-E_B}\,r_A^i\,p_A^j\,r_B^k\biggr\rangle\nonumber
              \\ &&-\frac{2}{15}\,\biggl\langle
              r_A^i\,r_B^j\,\frac{1}{H_A+H_B-E_A-E_B}\,(2\,r_A^2\,p_A^i\,-r_A^i\,\vec
              r_A\cdot\vec p_A)\,r_B^j\biggr\rangle \,,\label{32}\\
C_5^{(5)} &=& \frac{7}{6\,\pi}\,\langle r_A^2\rangle\,N_B \,.\label{33}
\end{eqnarray}
where $N_A$ and $N_B$ are the number of electrons in the atom $A$ and $B$ respectively.
If $\delta^{(2)} E_{CP}$ includes contributions from the atom $B$, then
coefficients $C_6^{(2)}, C_6^{(4)}$ and $C_5^{(5)}$ should include
corresponding terms obtained by the replacement $A\leftrightarrow B$. 

The $C^{(i)}_j$ coefficients allow one to obtain a convenient form
of the interaction potential in the whole region of the atomic distance $R$,
as long as these atoms do not overlap. The minimal version of this potential is
\begin{equation}
E = E^{(2)}(m\,\alpha\,R) + E_{CP}(m\,\alpha^2\,R) + 
m\,\alpha^2\,\frac{C_6^{(2)}}{(m\,\alpha\,R)^6} 
\label{34}
\end{equation}
and the most accurate version using present result is
\begin{eqnarray}
E &=& E^{(2)}(m\,\alpha\,R)+E^{(4)}(m\,\alpha\,R) + E^{(5)}(m\,\alpha\,R)+
E_{CP}(m\,\alpha^2\,R) + \delta^{(2)} E_{CP}(m\,\alpha^2\,R)
\nonumber \\ &&
+m\,\alpha^2\,\biggl[\frac{C_6^{(2)}}{(m\,\alpha\,R)^6} + 
\frac{C_8^{(2)}}{(m\,\alpha\,R)^8}\biggr]
+m\,\alpha^4\,\biggl[\frac{C_4^{(4)}}{(m\,\alpha\,R)^4} + 
\frac{C_6^{(4)}}{(m\,\alpha\,R)^6}\biggr]\nonumber \\ &&
+m\,\alpha^5\,\biggl[\frac{C_3^{(5)}}{(m\,\alpha\,R)^3} + 
\frac{C_5^{(5)}}{(m\,\alpha\,R)^5}\biggr]\,.
\label{35}
\end{eqnarray}
 
In summary, the purpose of this work was the derivation of a complete $\alpha^2$ correction
to the Casimir-Polder potential in order to obtain a more accurate description
of inter-atomic interactions in the region where the electron wave functions from
different atoms do not overlap. The obtained result can be used for
the precise calculation of the scattering length and highly excited
vibrational levels of light molecules. Particularly interesting is the helium
dimer which existence has been confirmed as recently as in 1994
\cite{he_dimer}. Its dissociation energy has an extremely small value of
1 mK, while the mean internuclear distance is as large as 50 \AA.
Its existence can be associated to the long range attraction between the 
helium monomers. Since the minor perturbations of the interaction potential
result in significant changes in the description of the nuclear motion,
the potential in the large range of inter-atomic distances with an accuracy of
the order of 1 mK is needed,
which is the magnitude of relativistic and QED effects \cite{cencek}.

\section*{ACKNOWLEDGMENTS}
\noindent I wish to acknowledge interesting discussions with Bogumi\l\
Jeziorski and Grzegorz \L ach.

\end{document}